\documentclass{aastex}          
\usepackage{spr-astr-addons}    
\usepackage{url}
\usepackage{bm}                 
\usepackage{widetext}           
\usepackage{delarray}

\newcommand{\beq}{\begin{equation}}
\newcommand{\eeq}{\end{equation}}
\newcommand{\bea}{\begin{eqnarray}}
\newcommand{\eea}{\end{eqnarray}}
\newcommand{\bew}{\begin{widetext}}
\newcommand{\eew}{\end{widetext}}
\newcommand{\mrm}[1]{\mathrm{#1}}
\newcommand{\mbf}[1]{\mathbf{#1}}

\newcommand{\tens}[1]{\boldsymbol{#1}}
\newcommand{\p}[2]{p^{#1}_{#2}}
\newcommand{\del}{\nabla}
\newcommand{\tr}{\mrm{Tr}\,}
\newcommand{\re}{\mrm{Re}\,}
\newcommand{\dsub}[1]{\partial_{#1}}
\newcommand{\abs}[1]{\lvert{#1}\rvert}

\begin{document}
%
\title{MaxEnt power spectrum estimation using the Fourier transform for irregularly sampled data applied to a record of stellar luminosity}

\shorttitle{MaxEnt power spectrum estimation}
\shortauthors{R. W. Johnson}

\author{Robert W. Johnson}
\affil{Alphawave Research, Atlanta, GA 30238, USA} 
\email{robjohnson@alphawaveresearch.com} 


\begin{abstract}
The principle of maximum entropy is applied to the spectral analysis of a data signal with general variance matrix and containing gaps in the record.  The role of the entropic regularizer is to prevent one from overestimating structure in the spectrum when faced with imperfect data.  Several arguments are presented suggesting that the arbitrary prefactor should not be introduced to the entropy term.  The introduction of that factor is not required when a continuous Poisson distribution is used for the amplitude coefficients.  We compare the formalism for when the variance of the data is known explicitly to that for when the variance is known only to lie in some finite range.  The result of including the entropic measure factor is to suggest a spectrum consistent with the variance of the data which has less structure than that given by the forward transform.  An application of the methodology to example data is demonstrated.
\end{abstract}




\section{Introduction}
\label{sec:intro}
The analysis of imperfect data is a common task in science.  Given a set of measurements sampled over time, one commonly uses the Fourier transform to estimate the power carried by the signal as a function of frequency.  The forward transform is commonly viewed as the best estimate of the amplitude and phase associated with basis functions of independent frequency; however, the indiscriminate use of the forward transform is not appropriate when the data are known to be subject to measurement error, and the problem of irregular sampling is often addressed by \textit{ad hoc} methods of varying subjectivity, such as interpolation~\citep{cenker:1991,malik:2005} or zero-padding~\citep{Boyd:1992243}.

Bayesian statistical inference is a well-established methodology for dealing with imperfect data~\citep{Bretthorst-1988,Sivia-1996,gregory-2005}.  The parameters of interest, here the amplitude and phase comprising the power spectral density, are related to the data through a model function which may be nonlinear.  When that function is invertible, its inverse is usually called the forward transform of the data, but the methodology applies as well to model functions which are not invertible.  The most likely values for the parameters are given by those which maximize their joint distribution, which takes into account both their likelihood as measured by the discrepancy between the model and the data and their possibly non-uniform prior distribution.  A non-uniform prior commonly represents the invariant Haar measure under a change of variables, and when the number of parameters exceeds the number of data, a prior based on entropic arguments is often employed.  These methods generally fall under the rubric of MaxEnt data analysis, as the optimum of the joint distribution minimizes the residual while simultaneously maximizing the entropy distribution.  Essentially, we are extending the Lomb-Scargle method~\citep{Lomb-447,Scargle-835} to incorporate the effect of the measurement errors on the estimate of the most likely amplitude and phase coefficients.

After a brief review of Bayesian data analysis, we investigate the details of its application to power spectral density estimation using Fourier basis functions.  The problem of missing data values for otherwise regular sampling is easily addressed by working with basis functions defined only at the measurement times.  We will compare the methodology for the case when the variance matrix of the data is known explicitly to that for the more likely occurrence of when the variance is known only to lie in some finite range with assumed independent measurements.  The primary result is that the entropic prior flattens the power spectrum relative to that produced by the forward transform, which maximizes solely the likelihood distribution.  We demonstrate an application of the method to a signal derived from stellar observation, and we close by discussing recent ideas for improvement of the method so that the variance of the coefficients may also be evaluated.

\section{Bayesian primer}
\label{sec:primer}
The mathematical language of Bayesian data analysis is that of conditional probability distribution functions~\citep{Durrett-1994,Sivia-1996}.  We notate ``the probability of $A$ given $B$ under conditions $C$'' as \beq
\mathrm{prob}(A \vert B ; C) \equiv p(A \vert_C B) \equiv \p{A}{B} \;,
\eeq dropping the conditioning statement $C$ when it is unchanging, but its presence is always implied.  The sum and product rules of probability theory give rise to the expressions for marginalization and Bayes' theorem, \bea
\p{A}{} &=& \int_{\{B\}} \p{A, B}{} \, \mrm{d}B \;, \\
\p{A}{B} \p{B}{} &=& \p{B}{A} \p{A}{} \;,
\eea where marginalization follows from the normalization requirement and Bayes' theorem follows from requiring logical consistency of the joint distribution $\p{A, B}{} = \p{B, A}{}$.  Translated to data analysis, Bayes' theorem relates the evidence given data $\mbf{y}$ for the parameters $\mbf{X}$ yielding model $\mbf{x} = \mbf{x}(\mbf{X})$, denoted $\p{\mbf{X}}{\mbf{y}}$, to the likelihood of the data given the parameters $\p{\mbf{y}}{\mbf{X}}$ times the prior distribution for the parameters $\p{\mbf{X}}{}$, \beq
\p{\mbf{X}}{\mbf{y}} \propto \p{\mbf{y}}{\mbf{X}} \p{\mbf{X}}{} \;,
\eeq where the constant of proportionality $\p{\mbf{y}}{}$ represents the chance of measuring the data which in practice is recovered from the normalization requirement of the joint distribution.

\begin{figure}[b]
\includegraphics[width=\columnwidth]{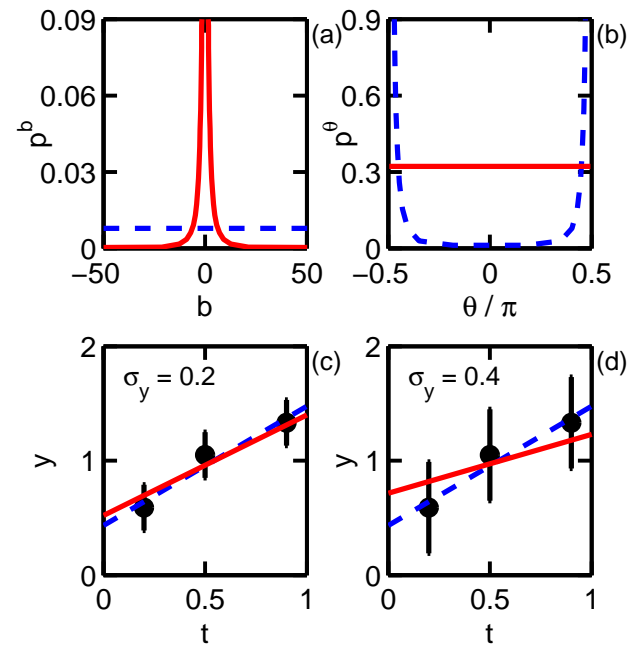}
\caption{Prior distributions (a) and (b) for the slope $b = \tan \theta$ of a linear fit.  The dashed line shows a prior uniform in $b$ while the solid line shows a prior uniform in $\theta$.  The model function $\mbf{x}_F$ (solid) has a slope with less magnitude than that of $\mbf{x}_R$ (dashed) which decreases as the variance increases from (c) to (d)}
\label{fig:A}       
\end{figure}

The essential feature of Bayesian data analysis which takes it beyond simple least-squares fitting is the use of a non-uniform prior in appropriate circumstances~\citep{dAgo-1998}.  The role of the prior is to prevent one from overestimating structure in the model not supported by imperfect data.  A prior which appears non-uniform in one's chosen variable generally represents a prior which is uniform under a change of variable to that with invariant Haar measure.  For example, consider fitting a two parameter model for a straight line $\mbf{x} = b\, \mbf{t} + a$ to a set of data $(\mbf{t}, \mbf{y})$ with finite $\sigma_y$ and $\sigma_t = 0$.  A maximum likelihood analysis (or least-squares for independent data) with a prior uniform on both $a$ and $b$ actually has a preferential bias for extreme values of the slope, as seen by transforming that distribution $\p{b}{} \propto 1$ to the variable for the angle $\tan \theta = b$.  That transformation is given by $\p{\theta}{} = \p{b}{} \, \abs{db / d\theta}$ such that $\p{\theta}{} \propto 1 + \tan^2 \theta$.  A prior which instead is uniform over the angle $\p{\theta}{} \propto 1$ leads to a Cauchy distribution for the slope $\p{b}{} \propto (1 + b^2)^{-1}$.  These priors are compared in Figure~\ref{fig:A} panels (a) and (b).  Centering the abscissa $\mbf{t} \rightarrow \mbf{t} - t_0$ for $t_0 \equiv \sum_d t_d \sigma_d^{-2} / \sum_d \sigma_d^{-2}$ yields independent estimates for the slope and intercept $\sigma^2_{ab} = 0$, and the prior for $a$ is uniform $p^a \propto 1$.  At $t_0$, the intercept $a$ is the estimate of the mean of $\mbf{y}$, through which the line of best fit must pass.

\begin{figure*}[t]
\centering
\includegraphics[]{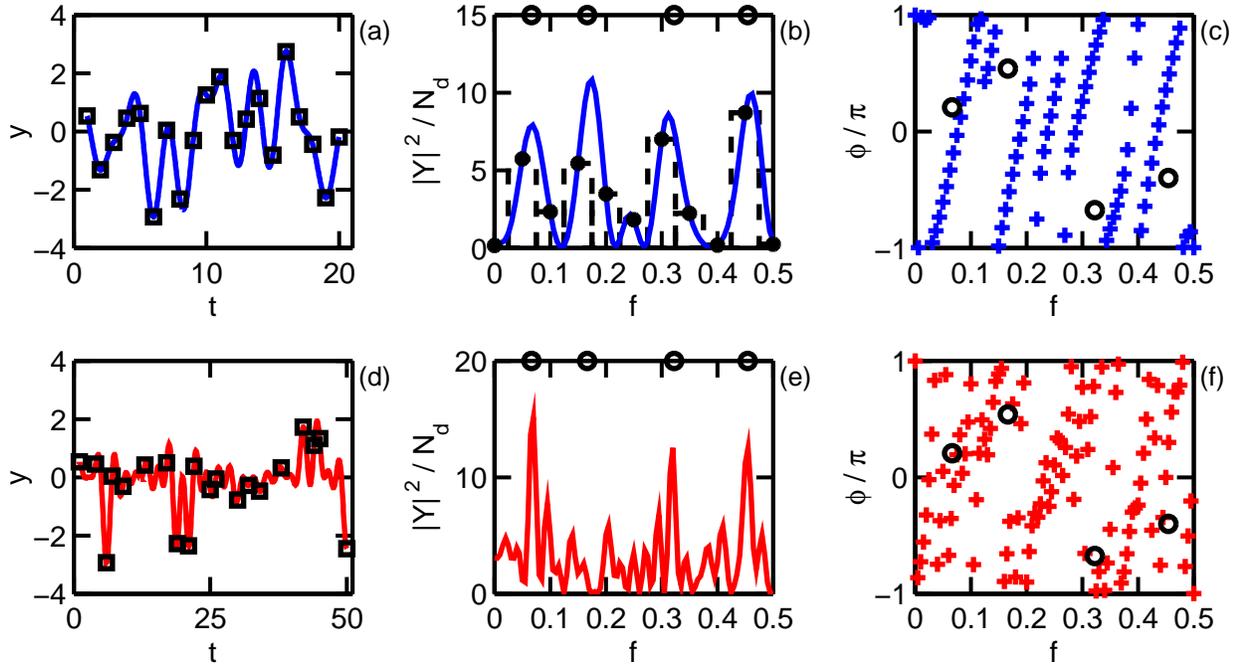}
\caption{A regularly sampled signal comprised of four sinusoids of unit amplitude, displayed as squares in (a), has the power spectrum shown in (b), with the signal frequencies indicated by circles at the top and the discrete Fourier transform periodogram marked by dots with the bin widths as dashed lines, and the phase spectrum is shown in (c) with the values of the signal phases circled.  When irregularly sampled with the same number of measurements (d), the power spectrum (e) is able (but not guaranteed) to resolve the signal components.  The phase spectrum (f) is not as clearly structured as for regular sampling.  The interpolated reconstructions displayed as lines in (a) and (d) reproduce the signal values at the measurement times}
\label{fig:B}       
\end{figure*}

The effect of a non-uniform prior is to shift the location of the solution with maximum evidence away from that given solely by the likelihood factor.  Writing the merit function $F = - \log \p{\mbf{X}}{\mbf{y}}$ as a sum of residual and prior terms $F = R + P$ and dropping the term for the normalization constant, a non-uniform prior $\del P \neq 0$ provides an optimal solution $\del F (\mbf{X}_F) = 0$ when the data have very little to say $\del R \approx 0$.  Returning to our example, let us consider a set of $N_d$ independent measurements with uniform variance and express the residual term $R = r^2 / 2 \sigma_y^2$ for $r^2 \equiv \sum_d (x_d - y_d)^2$.  As the data become less reliable $\sigma_y^2 \rightarrow \infty$, the prior term dominates the gradient and the estimate for $b$ tends to zero, indicating that the data may be characterized solely by their mean rather than exhibiting a linear relation to the abscissa.  We compare the model vectors $\mbf{x}_R$ and $\mbf{x}_F$ for a set of three data points at two values of $\sigma_y^2$ in Figure~\ref{fig:A} panels (c) and (d), where one can see how the slope diminishes as the variance increases.  The role of the prior is to prevent one from overestimating structure in the model when working with imperfect data by taking into account the measure factor for the chosen parametrization.  In that sense, Bayesian analysis provides the most conservative estimate consistent with the data.

\section{One-sided discrete CFT}
\label{sec:osdcft}
Let us now briefly review the theory of the continuous Fourier transform (CFT) in terms of its discrete application (dCFT).  Suppose there exists a signal $y_u(t_u)$, where the subscript $u$ reminds us that the data are given in terms of unit bearing quantities, sampled at regular intervals $t_u \equiv t \Delta_t$ for integer $t \in [1, N_t]$ and defining the unit of time $\Delta_t \equiv 1$, possibly with missing values $N_d < N_t$ such that $\mbf{t}$ contains only the measurement times and $\mbf{y}$ the corresponding values.  The critical frequency for aliasing is $f_c \equiv (2 \Delta_t)^{-1}$ and relates to the periodicity of the spectrum on an infinite frequency axis.  For a real signal $y_u \equiv y u_y$ the amplitude spectrum has conjugate symmetry about zero frequency, and so we can restrict consideration to the positive frequency axis $f_u \equiv f \Delta_f$ for integer $f \in [0, N_f]$ and $\Delta_f \equiv (2 N_f \Delta_t)^{-1}$.  The Fourier basis functions may be represented as a matrix $\tens{\Theta}$, where \beq
\Theta_{f t} = \sqrt{2} \exp (i 2 \pi f_u t_u) = \sqrt{2} \exp (i \pi f t / N_f) \;,
\eeq so that the forward transform $Y_f = \sum_t \Theta_{f t} y_t \Delta_t$ can be written as a matrix multiplication $\mbf{Y} = \tens{\Theta} \tens{D}_t \mbf{y}$, where $\tens{D}_t = \Delta_t \tens{I}$ is a diagonal matrix whose entries are all $\Delta_t$.  The factor $\sqrt{2}$ accounts for the one-sided nature of the transform, representing the response at negative frequencies which differs only by conjugation.  The signal energy defined by the sum of squared data values, \beq \label{eqn:Eysum}
E_y \equiv \sum_t y_t^2 \Delta_t = \mbf{y}^T \tens{D}_t \mbf{y} \;,
\eeq is equal to the spectral energy defined in terms of the transform coefficients $E_Y \equiv \mbf{Y}^\dagger \tens{D}_f \mbf{Y}$, where $\tens{D}_f$ is a diagonal matrix whose entries are $\Delta_f$ except for the first and last which are $\Delta_f / 2$.  The edge-most pixels corresponding to frequencies 0 and $N_f$ have a width only half that of the others when evaluating the integral over the frequency axis $\int_0^{1/2} df_u \rightarrow \sum_f \Delta_f$ in order for the limits to be strictly respected, in contrast to the usual conventions for forming a one-sided power spectral density (psd) from a two-sided Fourier transform which gives those pixels the same weighting as the others.  The units of the signal energy differ from those of physical energy by a factor of the load impedance, and the amplitude of the transform coefficients carries units of $u_Y = u_y \Delta_t$, as seen from the inverse transform $\mbf{y} = \re \tens{\Theta}^\dagger \tens{D}_f \mbf{Y}$, where $\tens{\Theta}^\dagger \equiv (\tens{\Theta}^*)^T$ is the conjugate transpose of the basis functions.

\begin{figure}[t]
\includegraphics[width=\columnwidth]{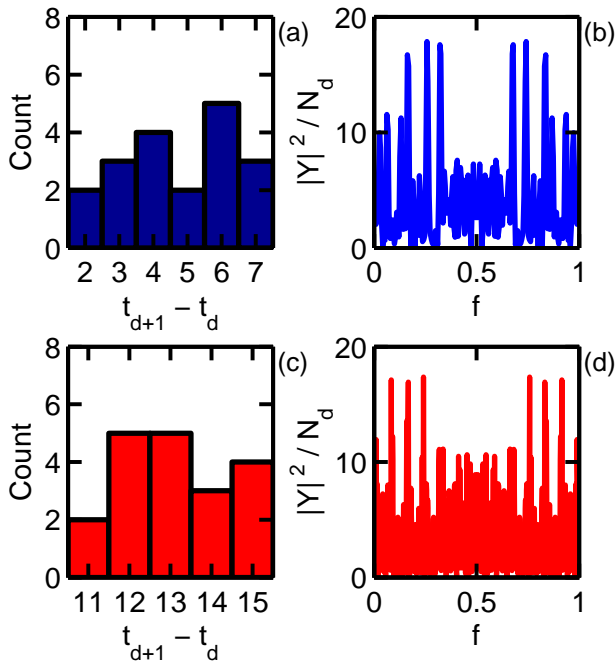}
\caption{The critical frequency $f_c$ for irregular sampling depends upon the greatest common factor of the measurement times.  In (a) and (c) are histograms of the inter-measurement period for two samplings of the same signal, and their corresponding power spectra in (b) and (d) are symmetric about the critical frequency}
\label{fig:C}       
\end{figure}

In order for the reconstruction to replicate the data (up to round-off errors), one must take a sufficient discretization of the frequency axis.  For regularly sampled data $N_d = N_t$, one has the requirement $N_f \geq \lceil N_t / 2$, and when $N_f$ is at its minimum the psd of the one-sided dCFT corresponds to that of the discrete Fourier transform periodogram~\citep{Press-1992} up to the factors of 2 for the edge-most pixels, as seen in Figure~\ref{fig:B} panels (a) through (c).  As $N_f$ increases, the resolution along the frequency axis improves so that the remainder of the CFT is evaluated.  For irregularly sampled data $N_d < N_t$, one requires $N_f \simeq N_t / 2$, but to achieve sufficient resolution it is better to take $N_f = 2 N_t$, as shown in Figure~\ref{fig:B} panels (d) through (f).  Virtually all cases of irregular sampling will correspond to the missing values problem once the greatest common factor of the measurement times is identified as $\Delta_t$, but if the individual measurement durations are not all equal then a suitably generalized $\tens{D}_t$ must be used.  To interpolate (or extrapolate) the data, one simply replaces $\mbf{t} \rightarrow \mbf{t}'$ in the inverse transform.  One caveat is that the inverse transform is required to agree with the data only at the measurement times, so that for increasing resolution along the frequency axis the interpolant is driven progressively towards the signal mean.

The identification of $f_c$, which is the lowest positive (nonzero) frequency whose basis function is entirely real over the measurement times, is confirmed by evaluating the psd over the domain $f_u \in [0, 1]$, as seen in Figure~\ref{fig:C}, where panels (a) and (c) are histograms of the inter-measurement period for two signals and the corresponding spectra in (b) and (d) are symmetric about $f_u = 1/2$, recalling $\Delta_t \equiv 1$.  There is no great mystery as to how an irregular sampling can resolve a frequency above the Nyquist limit for the corresponding regular case, as it is the Nyquist limit which must be defined in terms of the irregular sampling.  When implementing the continuous Fourier transform in a discrete setting, it is the sampling of the signal which induces the periodicity in the spectrum, while the finite signal duration causes side-lobes to appear in the point spread function~\citep{rwj:nova01}.  The location of the critical frequency $f_c$ must be known so that the normalization of the power spectrum can be evaluated over one (half) period of the frequency axis.

\begin{figure}[t]
\includegraphics[width=\columnwidth]{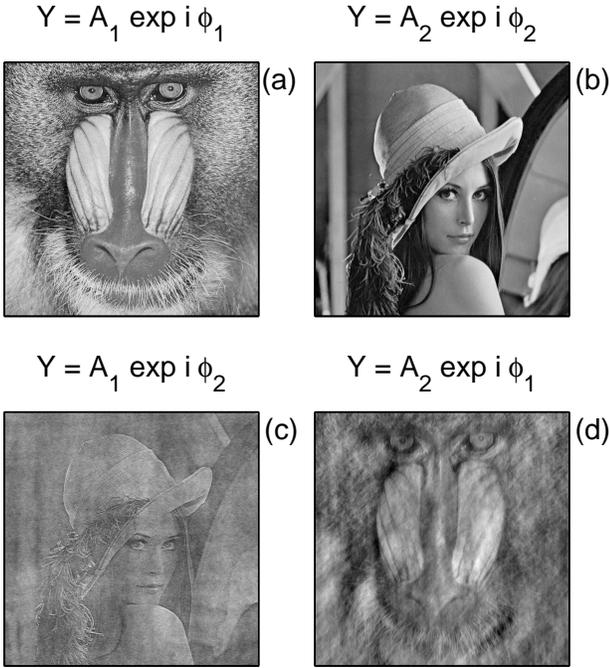}
\caption{The phase spectra of two images Mandrill (a) and Lena (b) may be exchanged to produce two new images (c) and (d) which resemble the image associated with the phase more than that associated with the amplitude}
\label{fig:D}       
\end{figure}

While the amplitude or power spectrum usually receives more attention from investigators, it is actually the phase spectrum which carries most of the information contained in the signal.  An amusing example of this phenomenon is the demonstration of how to make a mandrill look like a girl.  Taking the 2D Fourier transform of two standard test images shown in Figure~\ref{fig:D} panels (a) and (b) using a stock FFT algorithm, one can combine the amplitude spectrum of one image with the phase spectrum of the other to produce two new images from the inverse transform.  The resulting images will appear to the eye to resemble the original image associated with the phase used in the combination rather than the amplitude.  These two combinations of our test images are displayed in Figure~\ref{fig:D} panels (c) and (d), where indeed exchanging the phases has made the mandrill look like a girl and \textit{vice versa}.

\section{MaxEnt psd estimation for known data variance}
\label{sec:knownvar}
An important feature of Bayesian data analysis is that it allows one formally to address imperfect data as well as to incorporate a prior contribution to the evidence distribution.  Rather than assigning errors to the forward transform coefficients based on the data variance, the evidence for a set of coefficients is evaluated from the likelihood of measuring the signal given its variance, and the prior accounts for the measure factor for the chosen parametrization.  When there are many more parameters than data values [what \citet{Sivia-1996} calls a ``free-form'' model], an entropic prior is often used~\citep{Skilling-1989,BuckandMac:1992}; however, its usual expression is derived from an approximation to the Poisson distribution which introduces unnecessarily a factor for the conversion of units.  That factor is often said to be a Lagrange multiplier yet is treated as if it were a parameter of the model, which not a consistent approach.  With the obvious generalization of the discrete Poisson distribution to a continuum (described in the Appendix), the prior for the amplitude coefficients can be written without reference to an arbitrary unit factor.  In this section, we construct the merit function in terms of physically relevant quantities and show its reduction to the usual form as a consequence of normalization.  In its physically normalized form, this merit function can be related to those used in statistical physics and lattice gauge theory.

\subsection{Model, residual, and constraint}
\label{subsec:MRC}
Suppose that we are also given along with $\mbf{y}$ a variance matrix $\tens{V}$ whose entries are the covariance of the measurements $V_{jk} \equiv \sigma^2(y_j, y_k)$, from which we form the normalized weight matrix $\tens{W} \equiv \tens{V}^{-1} / \tr \tens{V}^{-1}$ such that $\tr \tens{W} = 1$.  One then defines the residual signal energy $R_E \equiv \mbf{r}^T \tens{W} \mbf{r} N_d \Delta_t$ in terms of the residual vector $\mbf{r} \equiv \mbf{x} - \mbf{y}$, the weight matrix $\tens{W}$, and the trace of the time metric $\tr \tens{D}_t = N_d \Delta_t$, (\emph{cf}. the expression for $E_y$ with $\mbf{y} \rightarrow \mbf{r}$ and $\tens{D}_t \rightarrow \tens{W} N_d \Delta_t$).  The residual vector ultimately is expressed in terms of the model parameters for amplitude and phase $X_f \equiv A_f \exp^i (\phi_f)$, where the notation $\exp^a(x) \equiv (\exp x)^a = e^{a x}$ is similar in spirit to its trigonometric counterpart, through the model function \bea \label{eqn:modlfcn}
\mbf{x} &\equiv& \re \tens{\Theta}^\dagger \tens{D}_f \mbf{X} \;, \\
 &=& \Delta_f \sqrt{2} \left[ \dfrac{1}{2} \left( A_0 + A_{N_f} \cos \pi \mbf{t} \right) \right. \nonumber \\
 & & \left. + \sum_{f=1}^{N_f-1} A_f \cos \left( \phi_f - \pi f \mbf{t} / N_f \right) \right] \;,
\eea with parameter domains of $A_0$ and $A_{N_f} \in [-\infty, \infty]$, $A_f \in [0, \infty]$, and $\phi_f$ periodic in $[-\pi, \pi]$.  The frequencies of 0 and $f_c$ must have a phase of 0 or $\pi$ because their basis functions are entirely real over $\mbf{t}$.  The assignment of uniform priors $\p{A}{} \propto 1$ and $\p{\phi}{} \propto 1$ to the amplitude and phase parameters corresponds to finding the maximum likelihood solution for the spectrum given the data and its variance.

The likelihood of a signal $\mbf{y}$ with errors described by a variance matrix $\tens{V}$ given its ``true'' value $\mbf{x}$ is written as a multivariate Gaussian $\p{\mbf{y}}{\mbf{x}, \tens{V}} \propto \exp^{-1}(\chi^2/2)$, where $\chi^2 \equiv \mbf{r}^T \tens{V}^{-1} \mbf{r}$.  The subscript $E$ on $R_E$ reminds us that it carries units equal to the signal energy and must be normalized before appearing in the argument of an exponential function.  In statistical physics~\citep{wannier-1969}, the normalization of the action appearing in the Maxwell distribution is given by the fluctuation energy of the system.  Here, let us write the normalization as $\beta_{1/E} R_E$, where $\beta_{1/E} \equiv 1 / k T$ is the inverse thermal energy for some generalized temperature $T$ describing the uncertainty to which the measurements are subject.  Rearranging factors, we have \bea
\beta_{1/E} R_E &=& ( 2 \beta_{1/E} N_d \Delta_t / \tr \tens{V}^{-1} ) ( \mbf{r}^T \tens{V}^{-1} \mbf{r} / 2 ) \\
 &\equiv& \beta R \;,
\eea where $R = \chi^2 / 2$.  We now argue that $\beta = 1$ as follows.  Substituting for the thermal energy, \beq
\beta = \dfrac{N_d \Delta_t / \tr \tens{V}^{-1}}{k T / 2} \;,
\eeq and recalling that $k T / 2$ equals the average fluctuation energy per quadratic degree of freedom, we see that the numerator and denominator are equal to $\bar{\sigma}^2 \Delta_t$, where $\bar{\sigma}^2$ is the reciprocal of the average of the eigenvalues of $\tens{V}^{-1}$.  The normalized residual $R$ is seen to be the ratio of the residual signal energy given by the discrepancy of the model to the thermal energy given by the variance of the data.  If one were to scale the residual by some arbitrary factor $\beta \neq 1$, that would be tantamount to saying that the experimentalists contributing the measurements have misrepresented the ratio of the units of their signal to its deviation.

By itself, the residual term $R$ is not sufficient to identify uniquely a maximum likelihood solution to the optimization problem.  The reason is because the model function Equation~(\ref{eqn:modlfcn}) is surjective but not injective.  What that means is that, for a sufficient discretization of the frequency axis, there exists a continuous family of coefficients $\{\mbf{X}_R\}$ that can produce a vanishing residual $R(\mbf{X}_R) = 0$.  While the forward transform $\mbf{Y}(\mbf{y})$ is one-to-one, the inverse transform $\mbf{x}(\mbf{X})$ such that $\mbf{x} = \mbf{y}$ is many-to-one; the forward transform coefficients are identified uniquely as the member of $\{\mbf{X}_R\}$ whose spectral energy equals the signal energy $E_\mbf{X} = E_\mbf{y}$.  In order that the maximum likelihood solution should equal the forward transform coefficients, the merit function must be supplemented with a term enforcing the constraint.

Using subscripts to indicate which terms are appearing in the merit function, the negative log likelihood of the data, given the constraint on the power spectrum \beq
C \equiv \sum_f A_f^2 \Delta_f / E_y - 1 \;,
\eeq is written $F_{RC} \equiv R + \lambda C$, where $\lambda$ is a Lagrange multiplier enforcing $C = 0$.  The constrained optimum of $R$ coincides with the unconstrained solution of $\del_{\lambda, \mbf{X}} F_{RC} = 0$, which is a saddle point in the space $(\lambda, \mbf{X})$.  The presence of the constraint makes the assignment of errors to the coefficients difficult; not only are there correlations between the amplitudes and phases but also a restriction on the allowed directions the variation in the amplitudes may take.  If one amplitude increases, some other must decrease so that the normalization condition is respected.  For these reasons, the results appearing henceforth for the psd are understood to be conditioned on satisfaction of the constraint $C$, and its variance in principle is recoverable from the Hessian of the merit function but not in a form which is practically useful for the chosen parametrization.  In a later section we will discuss some recent thoughts on improvements to the method so that an estimate for the variance of the model parameters may be obtained.

\subsection{Entropy and the Poisson distribution}
\label{subsec:entropy}

The entropic spectral energy for unnormalized distributions~\citep{Skilling-1989,BuckandMac:1992,Sivia-1996} is typically written \beq \label{eqn:SExpr}
S_E \equiv \sum_f \left[ A_f^2 - m_f - A_f^2 \log ( A_f^2 / m_f ) \right] \Delta_f \;,
\eeq where the sum over $f$ is understood to take into account the frequency metric $\tens{D}_f$.  The factor $m_f$ represents the default model, which in this case is a flat spectrum $m_f \equiv m$ given by the Lebesgue measure with the same energy as the signal, $m = 2 E_y \Delta_t$ such that $m \int_0^{1/2} df_u = E_y$.  To evaluate the terms of $S_E$ when $A_f \rightarrow 0$, one needs to write $0 \log 0 = \log 0^0 = 0$.

To normalize the Lebesgue measure, one divides the entropic spectral energy by the signal energy, letting us write the normalized entropy $S \equiv S_E / E_y$ as \beq
S = \sum_f \left\lbrace A_f^2 \left[ 1 - \log ( A_f^2 / 2 E_y \Delta_t ) \right] \Delta_f / E_y \right\rbrace - 1 \;. 
\eeq  Under the condition $\sum_f A_f^2 \Delta_f = E_y$ one could manipulate the terms evaluating to a constant, reducing the entropy to the familiar expression $-S_C = \sum_f p_f \log p_f$ for $p_f = A_f^2 / 2 N_f E_y \Delta_t$ and respecting the pixel measure, but the numerical optimization is over the entire space of $\mbf{X}$ so that only the explicit constant may be dropped for satisfactory convergence of the algorithm.

\begin{figure}[t]
\includegraphics[width=\columnwidth]{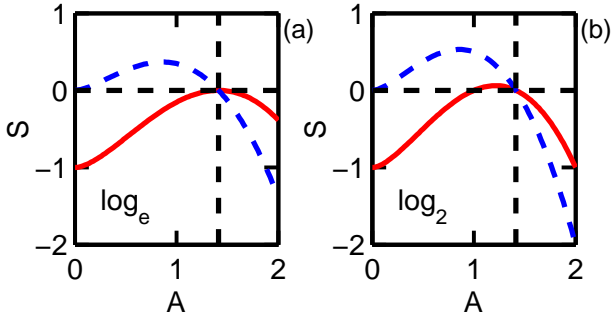}
\caption{Comparison of the entropy expressions $S_E$ (solid) and $S_C$ (dashed) for $E_y = 1$ and $N_f = 1$ using logarithm base $e$ in (a) and base 2 in (b), with $A = \sqrt{2}$ and $S = 0$ indicated by dashed lines}
\label{fig:J}       
\end{figure}

The quantity of physical relevance is the negentropy, which is the difference between the maximum attainable entropy and that of any particular configuration and so by definition not negative.  The equivalent requirement for the entropy expression is that it be non-positive.  For the following examples, let us suppose $E_y = 1$, and consider first a single frequency $N_f = 1$ whose pixel spans the Nyquist interval $\Delta_f = (2 \Delta_t)^{-1}$ such that $m = 2$.  In Figure~\ref{fig:J} panel (a) we compare $S_E$ and $S_C$ for this case, and we see that they do indeed agree at the location which satisfies the constraint, $A = m^{1/2}$, where both expressions equal zero.  The reduced entropy $S_C$, however, takes both positive and negative values over the unconstrained amplitude axis.  Replacing the base $e$ of the logarithm with base 2 such that $S_C$ equals the Shannon entropy, as shown in panel (b), the expressions again agree at the value of the constraint, but now the corresponding $S_E$ as well takes on positive values.  We are thus led to believe that the physical form of the negentropy expression is that given by $S_E$ in Equation~(\ref{eqn:SExpr}) normalized by the Lebesgue measure and with the natural logarithm.

\begin{figure}[t]
\includegraphics[width=\columnwidth]{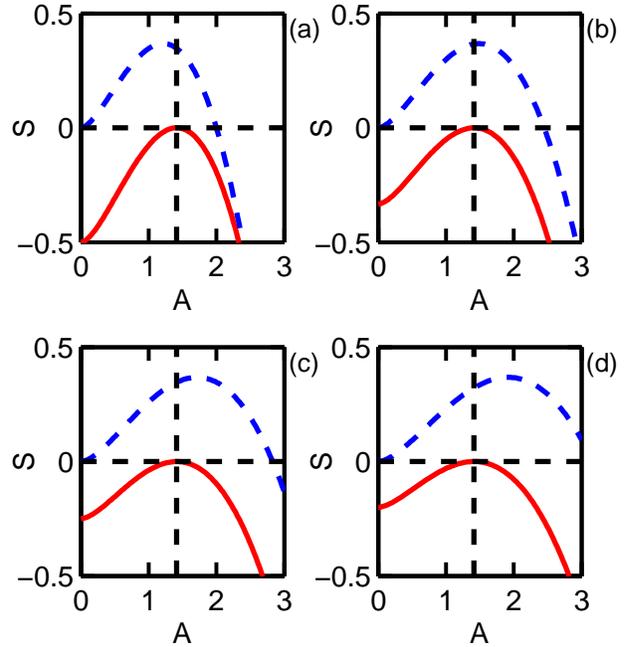}
\caption{Comparison of the entropy expressions $S_E$ (solid) and $S_C$ (dashed) for $E_y = 1$ and $N_f$ taking the value 2 in (a), 3 in (b), 4 in (c), and 5 in (d), with the maximum of $S_E$ at $A = \sqrt{2}$ and $S = 0$ indicated by dashed lines}
\label{fig:K}       
\end{figure}

Let us now compare the expressions for $S_E$ and $S_C$ as the number of frequencies spanning the Nyquist interval increases.  In Figure~\ref{fig:K} we show those expressions as $N_f$ goes from 2 to 5.  (We are assuming here a frequency grid dual to the one used elsewhere so that there are no edge effects.)  The constraint $C$ is now satisfied not by a single value but by $N_f$ values for $A$ with the required sum of squares.  The most obvious difference in behavior is that the maximum of $S_C$ does not remain fixed at $m^{1/2}$ the way that it does for $S_E$.  The presence of both positive and negative values for $S_C$ has a remarkable impact when one follows the mainstream approach of multiplying the entropy by a factor $\alpha$ claimed to play the role of a Lagrange multiplier in the merit function $F_{RC\alpha} \equiv R + \lambda C - \alpha S$ for $S = S_C$.  Under enforcement of the constraint $S_C = 0$ for nontrivial amplitudes $\mbf{A} \neq 0$, contributions with positive entropy must be balanced by contributions with negative entropy, so that the amplitude spectrum is drawn by the residual in either direction away from its nontrivial value where each contribution is zero.  The effect is to induce a spikiness to the spectrum, where relatively few large amplitudes with negative entropy (which is unbounded) are balanced by many small amplitudes with positive entropy (which is bounded).  While resolution enhancement is often a desired goal (which implies consideration of some point spread function), the use of a term $\alpha S_C$ is not the correct way to go about it.  When one uses the normalized negentropy $F_{RC\alpha}$ for $S = S_E / E_y$, the maximum of entropy coincides with the only values which can satisfy the constraint $S_E = 0$, namely the coefficients given by the default model $A_f \equiv m^{1/2}$.

The use of $\alpha$ as a Lagrange multiplier brings up various questions regarding the stopping criterion for the algorithm~\citep{Bryan-1990,strauss:1993,mackay:1999}.  Many practitioners use some variation of the method described by~\citet{Sivia-1996} in which $\alpha$ is treated as a parameter of the model; however, that approach is not consistent with the interpretation of a Lagrange multiplier whose sole purpose is to equate the norms of the gradient vectors for the residual and the constraint.  The stationary points of the Lagrangian merit function describe locations satisfying the constraint where the residual changes only in directions which are forbidden.  Similarly to the remarks concerning $\beta$, the use of $\alpha \neq 1$ is tantamount to an arbitrary rescaling of units between the entropic and signal energies $S_E$ and $E_y$.  By writing $F_{RSC} \equiv R - S + \lambda C$, the MaxEnt solution $\mbf{X}_{F}$ is carried smoothly from the forward transform coefficients $\mbf{X}_{RC} = \mbf{Y}$ for $\bar{\sigma}^2 \rightarrow 0$ to the coefficients with maximum entropy given by the Lebesgue measure as the mean magnitude of the data variance increases, $\mbf{X}_{F} \rightarrow \mbf{X}_{SC}$ as $\bar{\sigma}^2 \rightarrow \infty$.

Returning to the literature~\citep{Skilling-1989,BuckandMac:1992}, let us reexamine the arguments leading to the entropy expression found in Equation~(\ref{eqn:SExpr}).  When invoking the hypothetical troop of monkeys, one supposes not only that the image, here the power spectrum, is comprised of discrete pixels $\Delta_f$ but also that the pixel values are themselves described discretely in terms of some presumably small quantum of image, here power, such that $A_f^2 = \epsilon \, n_f$ for integer $n_f$, in order to write the measure as a product of discrete Poisson distributions $p(n_f|\nu_f)$ with parameters $\nu_f \equiv \nu$, taken here to be uniform.  Considering a single pixel $\p{n}{\nu}$, the requirement that $\epsilon$ be small so that $n$ is large allows the Stirling approximation to the factorial, \bea
\p{n}{\nu} &=& e^{-\nu} \nu^n / n! \\
 &\approx& (2 \pi n)^{-1/2} \exp ( n - \nu - n \log n / \nu ) \;,
\eea and transforming variables to those for amplitude $a^2 = n$ such that $dn / da = 2 a$ lets one write \bea
\p{a}{\nu} &=& \p{n}{\nu} \, \abs{dn / da} \;, \\
 &\approx& (2 / \pi)^{1/2} \exp ( a^2 - \nu - a^2 \log a^2 / \nu ) \;.
\eea  The inverse of the quantum $\epsilon^{-1} \equiv \alpha$ is then identified as the prefactor in the expression $\alpha S_E = \sum_f [ a_f^2 - \nu - a_f^2 \log ( a_f^2 / \nu ) ] \Delta_f$.  The motivation for the introduction of $\alpha$ hinges entirely on the discrete form of the Poisson distribution.  Its definition here as a unit factor for quantization is not consistent with its interpretation previously as a Lagrange multiplier.

\begin{figure}[t]
\includegraphics[width=\columnwidth]{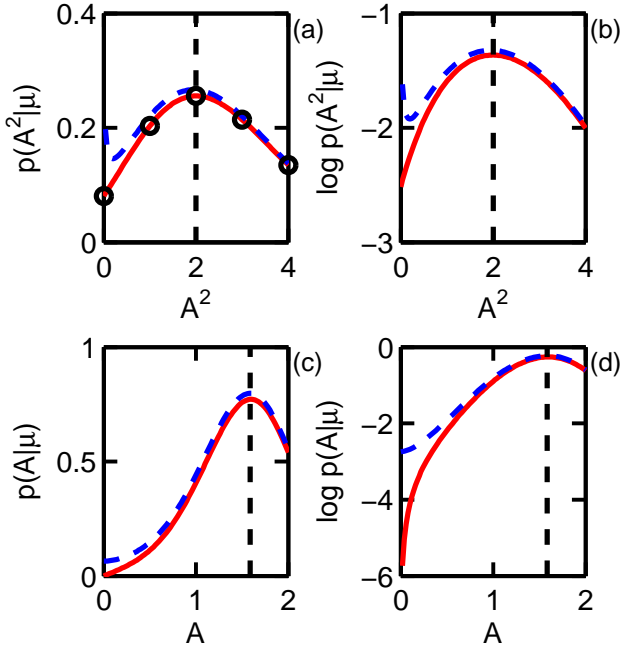}
\caption{Comparison of the continuum Poisson distribution (solid) with its Stirling approximation (dashed) for $m=2$, with the peaks at $A^2_m$ and $A_\mu$ indicated by dashed lines and the corresponding values of the discrete Poisson distribution circled in (a)}
\label{fig:L}       
\end{figure}

We propose that there is no need to introduce $\alpha$ when one considers the extension of the Poisson distribution to the continuum using the obvious substitution $n! \rightarrow \Gamma (n+1)$, where now $n = A^2$ and the parameter in the given units is $\mu$.  In the Appendix we discuss the use of this expression as a probability density function.  For a single pixel, the measure of the power distribution is given by \beq
p(A^2 | \mu) = e^{-\mu} \mu^{A^2} / \Gamma (A^2+1) \;,
\eeq which if maximized at a value $A_m^2 = m$ has a parameter $\mu_m = \exp [\Lambda_1 (m+1)]$, using the notation $\Lambda_k (r) \equiv (\partial_r)^k \log \Gamma (r)$ for the polygamma functions with real argument and integer $k \geq 0$.  In terms of amplitude, the distribution is \beq
p(A | \mu) \propto \abs{A} \, p(A^2 | \mu) \;,
\eeq which has its peak not at $A_m$ but at $A_\mu = \mu^{1/2}$.  Using the notation $q \equiv - \log p$, one can write \bea
q(A^2 | \mu) &=& \mu - A^2 \log \mu + \Lambda_0 (A^2+1) \;, \label{eqn:qA2} \\
q(A | \mu) &=& q(A^2 | \mu) - \dfrac{1}{2} \log A^2 \label{eqn:qA} \;,
\eea having dropped a factor of 2.  Under the Stirling approximation, \beq
\Lambda_0 (A^2+1) \approx A^2 \log A^2 - A^2 + \dfrac{1}{2} \log (2 \pi A^2) \;,
\eeq the expressions for $q$ are \bea
q(A^2 | \mu) &\approx& \mu - A^2 + A^2 \log (A^2 / \mu) + \dfrac{1}{2} \log A^2  \;, \\
q(A | \mu) &\approx& \mu - A^2 + A^2 \log (A^2 / \mu) \;,
\eea having dropped numerical terms.  The full expressions for $p$ and $q$ are compared for $m=2$ in Figure~\ref{fig:L}, and in panel (a) the corresponding values of the discrete Poisson distribution are circled.  Using Equations~(\ref{eqn:qA2}) and (\ref{eqn:qA}), we can now write the prior measure for the amplitudes in terms of its contribution $P$ to the merit function $F_{RPC} \equiv R + P + \lambda C$ as \beq \label{eqn:Pprior}
P = \sum_f [ \Lambda_0 (A_f^2+1) - \dfrac{1}{2} \log A_f^2 - A_f^2 \log \mu ] \Delta_f / E_y \;,
\eeq taking into account the pixel measure $\tens{D}_f$ and dropping the constant term proportional to $\mu$.  The parameters $A_0$ and $A_{N_f}$ have twice the range but only half the pixel width, so their prior normalization is consistent with the others.

\subsection{Finding the solution}
\label{subsec:soln}
To solve the optimization problem for $F = F_{RPC}$, one needs the gradient $\mbf{G} \equiv \del F$ and the Hessian $\tens{H} \equiv \del^T \del F$, with $\del^T$ a column vector of derivatives $\del^T \equiv [ \dsub{\lambda}, \dsub{\mbf{X}} ]^T$ for $\dsub{\mbf{X}}$ (a row vector) the covariant gradient in $\mbf{X}$, written so that matrix multiplication is embodied in the notation.  The gradient of the residual vector $\dsub{\mbf{X}} \mbf{r} = \dsub{\mbf{X}} (\mbf{x} - \mbf{y})$ is a matrix with a column index for the parameters $\mbf{X}$ and a row index for the measurement times $\mbf{t}$.  The Hessian operator has the dyadic form \beq
\del^T \del \equiv \left[ \begin{array}{c} \dsub{\lambda} \\ \dsub{\mbf{X}}^T \end{array} \right] \left[ \begin{array}{cc} \dsub{\lambda} & \dsub{\mbf{X}} \end{array} \right]  =  \left[ \begin{array}{cc} \dsub{\lambda}^2 & \dsub{\lambda} \dsub{\mbf{X}} \\ \dsub{\lambda} \dsub{\mbf{X}}^T & \dsub{\mbf{X}}^T \dsub{\mbf{X}} \end{array} \right] \;,
\eeq  and the $\lambda$ dependence is $\dsub{\lambda} F = C$, $\dsub{\lambda}^2 F = 0$, and $\dsub{\lambda} \dsub{\mbf{X}} F = \dsub{\mbf{X}} C$.  The residual term has contributions \bea
\dsub{\mbf{X}} R &=& \mbf{r}^T \tens{V}^{-1} \left( \dsub{\mbf{X}} \mbf{r} \right) \;, \\
\dsub{\mbf{X}}^T \dsub{\mbf{X}} R &=& \left( \dsub{\mbf{X}} \mbf{r} \right)^T \tens{V}^{-1} \left( \dsub{\mbf{X}} \mbf{r} \right) + \mbf{r}^T \tens{V}^{-1} \left( \dsub{\mbf{X}}^T \dsub{\mbf{X}} \mbf{r} \right) \;,
\eea where $\left( \dsub{\mbf{X}}^T \dsub{\mbf{X}} \mbf{r} \right)$ is contracted along its time index with $\mbf{r}^T \tens{V}^{-1}$.  Ordering the parameter vector using the notation $\mbf{X}^T \equiv [A_0, A_{N_f}, [A_f, \phi_f]_{f=1}^{N_f-1}]$ and letting $\mbf{u}_f \equiv \pi f \mbf{t} / N_f$, we can write \bew \bea
( \dsub{\mbf{X}} \mbf{r} )^T &=& \dfrac{\Delta_f}{\sqrt{2}} \left[ \begin{array}{c} 1 \\ \cos \left( \pi \mbf{t}^T \right) \\ \begin{array}[c] [{c}{]_f} 2 \cos \left( \phi_f - \mbf{u}_f^T \right) \\ - 2 A_f \sin \left( \phi_f - \mbf{u}_f^T \right) \end{array} \end{array} \right] \;, \\
\dsub{\mbf{X}}^T \dsub{\mbf{X}} \mbf{r} &=& \dfrac{\Delta_f}{\sqrt{2}} \left[ {\setlength\arraycolsep{0.2em} \begin{array}{cccc} 0 & 0 & 0 & 0 \\ 0 & 0 & 0 & 0 \\ \begin{array}{l} 0 \\ 0 \end{array} & \begin{array}{l} 0 \\ 0 \end{array}  & {\setlength\arraycolsep{0.0em} \begin{array}[c] [{c}. 0 \\ - 2 \sin \left( \phi_f - \mbf{u}_f \right) \end{array} } & {\setlength\arraycolsep{0.0em} \begin{array}[c] .{c}{]_f} - 2 \sin \left( \phi_f - \mbf{u}_f \right) \\ - 2 A_f \cos \left( \phi_f - \mbf{u}_f \right) \end{array} } \end{array} } \right] \;,
\eea and for $\Psi_\pm(A) \equiv \Lambda_1 (A^2) \pm (2 A^2)^{-1} - \log \mu + \lambda$, one has \bea
\dsub{\mbf{X}}^T (P + \lambda C) &=& \dfrac{\Delta_f}{E_y} \left[ \begin{array}{c} A_0 \Psi_{+}(A_0) \\ A_{N_f} \Psi_{+}(A_{N_f}) \\ \begin{array}[c] [{c}{]_f} 2 A_f \Psi_{+}(A_f) \\ 0 \end{array} \end{array} \right] \;, \\
\dsub{\mbf{X}}^T \dsub{\mbf{X}} (P + \lambda C) &=& 
 \dfrac{\Delta_f}{E_y} \left[ {\setlength\arraycolsep{0.2em} \begin{array}{cccc} 2 A_0^2 \Lambda_2 (A_0^2) + \Psi_{-}(A_0) & 0 & 0 & 0 \\ 0 & 2 A_{N_f}^2 \Lambda_2 (A_{N_f}^2) + \Psi_{-}(A_{N_f}) & 0 & 0 \\ \begin{array}{c} 0 \\ 0 \end{array} & \begin{array}{c} 0 \\ 0 \end{array} & \begin{array}[c] [{c}. 4 A_f^2 \Lambda_2 (A_f^2) + 2 \Psi_{-}(A_f) \\ 0 \end{array} & \begin{array}[c] .{r}{]_f} \quad 0 \\ 0 \end{array} \end{array} } \right] \;.
\eea \eew  

\begin{figure}[t]
\includegraphics[width=\columnwidth]{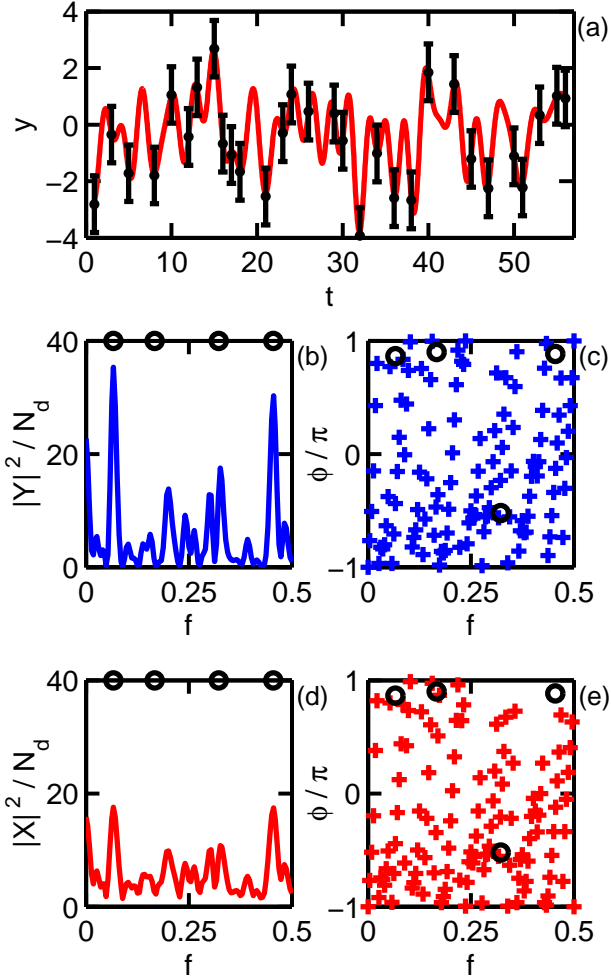}
\caption{The irregularly sampled signal is displayed in (a) as dots with error-bars.  The maximum likelihood psd in (b) is given by the forward dCFT, and its phases are shown in (c).  The maximum evidence psd in (d) accounts for the variance of the data by bringing the psd closer to a flat spectrum, and the phases in (e) have varied slightly.  The reconstructed signal is shown in (a) as a solid line}
\label{fig:E}       
\end{figure}

With these evaluations, the solution with maximum evidence $\mbf{X}_F$ may be found using commonly available numerical optimization routines~\citep{Press-1992}.  Let us compare the amplitude spectra for $\mbf{Y}$ and $\mbf{X}_F$ given a signal with unit variance and missing values.  The simulated signal displayed in Figure~\ref{fig:E}~(a) is comprised of four sinusoids of unit amplitude and Gaussian noise of unit variance and has the forward transform power spectral density $\vert Y_f \vert^2$ shown in (b) and the phase spectrum shown in (c), where two of the four signal components have been well resolved for this particular sampling.  The effect of including the variance is seen in Figure~\ref{fig:E}~(d), where the peaks have been reduced to a level not far above the noise, and in (e), where the phases have adjusted slightly.  Recalling that the psd is proportional to a probability distribution, we see that the evidence for the signal components is reduced by a factor of nearly 2 compared to their likelihood when the signal variance is on the order of the signal magnitude squared.  Note that the signal energy of the reconstruction is generally less than the original signal energy, $E_x < E_y$, indicating that phase cancellations occur in the spectrum with maximum evidence.

The effect of the non-uniform prior for the amplitudes $A_f$ has been to draw the power spectrum towards the default model given by a flat spectrum.  The amount by which the spectrum is flattened depends upon the magnitude of the variance of the data.  The phase parameters $\phi_f$ are given a uniform prior so that they are free to adjust as needed to bring the model vector $\mbf{x}$ as close as possible to the data vector $\mbf{y}$.  In practice, the phases are not bound to their principle branch so that the algorithm does not get hung up on a branch cut; they are simply reduced to the principle branch after the optimization.  The maximum evidence solution is more conservative than that given by the forward transform, in that less structure is assigned to the power distribution.  Features which persist in the spectrum are the most likely to be of significance.

\section{MaxEnt psd estimation for unknown data variance}
\label{sec:unknownvar}
\begin{figure}[t]
\includegraphics[width=\columnwidth]{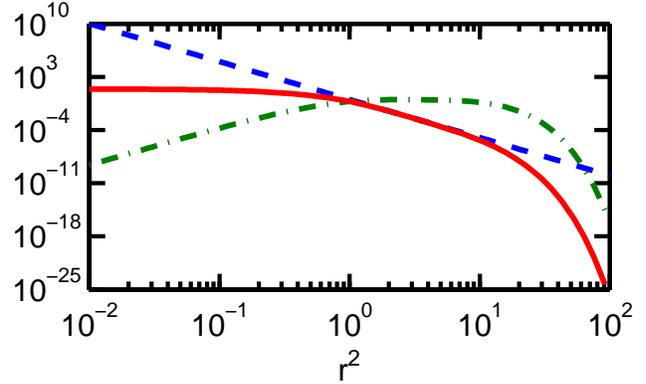}
\caption{The likelihood distribution (solid) for data with unknown variance is proportional to the product of two factors, $r^{-2a}$ (dashed) and $\Delta_\Gamma$ (dash-dot), here normalized by $\Gamma (a)$ for $a=5$ and $\sigma^2 \in [0.1, 1]$}
\label{fig:F}       
\end{figure}

Let us now look at how the methodology changes when the variance of the data is known only to lie in some finite range with assumed independent measurements.  The expressions for $P$ and $\lambda C$ remain the same, so we will focus on the changes to the residual term $R$.  It will prove convenient to work with the squared norm of the residual vector $r^2$, so that $\del r^2 / 2 =  \mbf{r}^T  (\del \mbf{r})$ and $\del^T \del r^2 / 2 = ( \del \mbf{r} )^T ( \del \mbf{r} ) + \mbf{r}^T  (\del^T \del \mbf{r})$.  The likelihood distribution is now expressed as an integral over the nuisance parameter for the data deviation $\sigma$, \beq
\int_{\sigma_0}^{\sigma_1} \sigma^{- N_d - 1} \exp \left( \dfrac{-r^2}{2 \sigma^2} \right) d\sigma \propto ( r^2 )^{- N_d / 2} \Delta_\Gamma \;,
\eeq where the integrand includes the Jeffreys prior $1 / \sigma$ as well as the normalization of the Gaussian and $\Delta_\Gamma \equiv \Gamma (N_d/2, r^2/2\sigma_1^2) - \Gamma (N_d/2, r^2/2\sigma_0^2)$ in terms of the upper incomplete gamma function $\Gamma (a, z) \equiv \int_z^\infty e^{-u} u^{a-1} du$.  In the limit of $\sigma_0 \rightarrow 0$ and $\sigma_1 \rightarrow \infty$, one has $\Delta_\Gamma \rightarrow \Gamma (a)$ which is absorbed by the normalization such that the minimum of $R$ is given by $r^2 \rightarrow 0$, but for a finite range $\sigma \in [\sigma_0, \sigma_1]$, the divergence of the factor $r^{-2a}$ is canceled by its reciprocal appearing in the continued fraction expression of $\Gamma (a, z) = \exp (-z) z^a (z + \ldots)^{-1}$ so that the likelihood is effectively constant for $r^2$ below a certain threshold and also reduced significantly for large values of $r^2$, as seen in Figure~\ref{fig:F}.

\begin{figure}[t]
\includegraphics[width=\columnwidth]{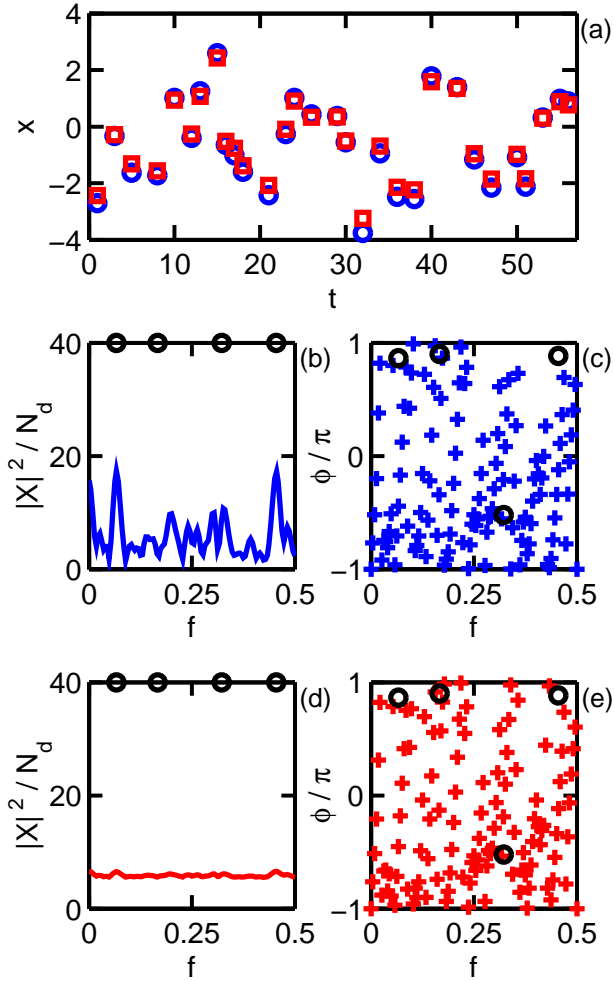}
\caption{The MaxEnt psd for unknown data variance in the range $\sigma_a \in [1, 10]$ in (b) and (c) is exceedingly similar to the psd for known variance $\sigma = 1$.  When the data is of poor quality, $\sigma_b \in [10, 100]$, the MaxEnt psd in (d) in (e) is very close to the default model of a uniform spectrum.  The reconstructions in (a), $\circ$ for $\sigma_a$ and $\Box$ for $\sigma_b$, are drawn closer to the signal mean as the lower bound on the data variance increases}
\label{fig:G}       
\end{figure}

Considering the same signal as in the previous section, let us first suppose that $\sigma_a \in [1, 10]$.  The resulting MaxEnt psd is shown in Figure~\ref{fig:G}~(b), and we see that it is nearly identical to that produced by the previous analysis which assumed a unit variance.  When the variance of the data is known only to lie in some finite range, the evidence is dominated by the contribution from the lower bound on the data variance.  Let us now suppose an extreme case of $\sigma_b \in [10, 100]$, where the variance exceeds the magnitude of the signal.  The psd in this case, Figure~\ref{fig:G}~(d), is now very close to the uniform distribution given by the default model of the prior term---the previously resolved signal components register barely a ripple on an otherwise flat power spectrum.  Probability theory has prevented us from overestimating structure in the spectrum when the quality of the data is suspect.  As the lower bound on the variance increases, the reconstruction from the model function grows closer to the mean of the data, as seen in Figure~\ref{fig:G}~(a), such that the reconstructed signal's energy decreases from the value given by the original data, $E_b < E_a < E_y$, indicating that phase cancellations become more prominent in the spectrum.

\section{Application to a record of stellar luminosity}
\label{sec:realdata}
Let us now apply the MaxEnt methodology to some real data.  The signal chosen here is a record of luminosity~\citep{henden:aavso} for the star VCas dating from the beginning of January, 2010, to the end of January, 2011.  Some measurements are given with a temporal resolution of seconds, but to keep the analysis tractable we assign the data to a daily time axis.  When more than one measurement falls on the same day (a rare occurrence) their mean and its variance are used.  To reveal the low frequency content, the arithmetic mean is subtracted before conducting the spectral analysis.  The choice of the arithmetic rather than the weighted mean is made because it is the arithmetic mean which appears in the 0 frequency bin of the forward transform.

\begin{figure}[t]
\includegraphics[width=\columnwidth]{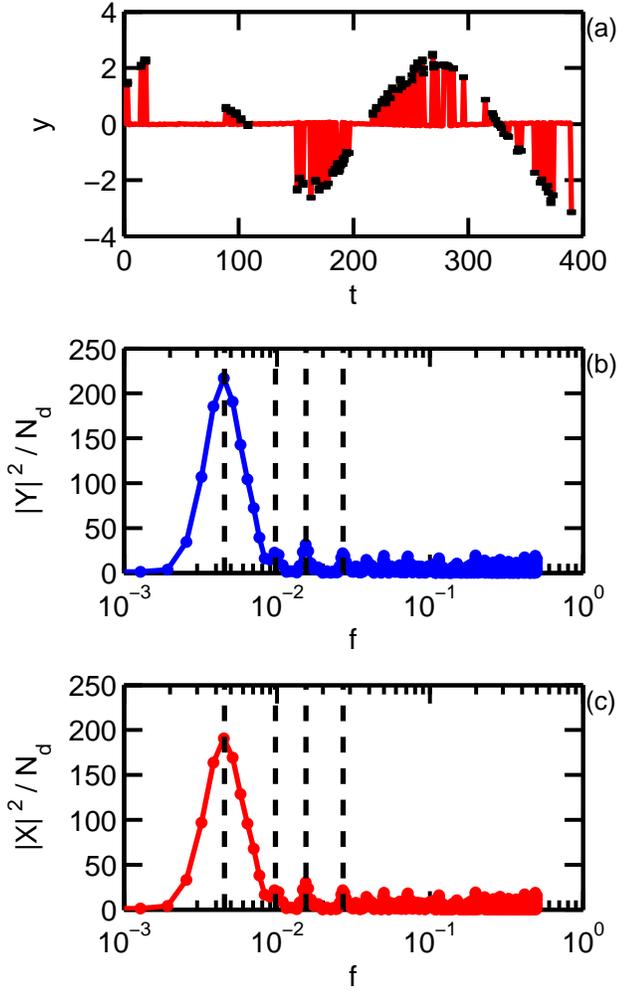}
\caption{The MaxEnt reconstruction of the VCas luminosity signal in (a) is driven to the signal mean between the measurement times by the oversampling of the frequency axis.  The time axis is given in units of days, and the frequency axis in units of cycles per day.  The forward transform psd in (b) displays low frequency structure which becomes more apparent on a logarithmic frequency axis, where the locations of the four lowest frequency peaks are indicated by dashed lines.  The MaxEnt psd in (c) is drawn only slightly towards a flat spectrum because of the small magnitude of the data variance.  The phase plots have been suppressed as little variation is seen between the forward transform and MaxEnt spectra}
\label{fig:H}       
\end{figure}

The data span $N_t = 390$ days, as shown in Figure~\ref{fig:H}~(a), and so a $N_f = 780$ point frequency axis is used.  The error bars are hard to see because they are small.  The MaxEnt reconstruction also shown in panel (a) is driven towards the signal mean because of the oversampling of the frequency axis.  The forward transform psd displayed in panel (b) shows a prominent low frequency peak as well as a few others with a period greater than 30 days.  The transform is evaluated on a linear frequency axis but displayed on a logarithmic axis so that the low frequency region is more easily observed.  The MaxEnt psd shown in (c) has not changed much, as expected from the small magnitude of the errors, but by incorporating the data variance and the amplitude measure, it provides a more conservative estimate of the spectral density.  The phase plots have been suppressed as very little variation is seen between the algorithms for this data set.

The locations of the four lowest frequency peaks are given in Table~\ref{tab:I}, as is their ratio to the lowest frequency.  These frequency peaks do not appear to be in a harmonic relation with the fundamental frequency of the signal.  Note that the MaxEnt algorithm does not alter the locations of the peaks but does broaden their widths so that the variance of a frequency estimate increases with the noise level of the data.  For this particular data set, a single or few frequency model~\citep{Bretthorst-1988} might be more appropriate; however, this signal was chosen simply as an example of the type of data to which the MaxEnt psd algorithm for irregular sampling is applicable.

\begin{table}[t]
\caption{Peak locations in the power spectral density of the VCas data in units of days}
\label{tab:I}       
\begin{tabular}{crrrr}
\tableline  
 & first & second & third & fourth \\
\tableline  
period & 220.82 & 102.11 & 64.67 & 36.98 \\
freq. & 0.004529 & 0.009793 & 0.015463 & 0.027038 \\
ratio & 1 & 2.163 & 3.415 & 5.971 \\
\tableline 
\end{tabular}
\end{table}

\section{Variance of the psd}
\label{sec:varest}
In this section we present some recent thoughts on how the methodology might be improved so that an estimate of the variance of the model parameters may be obtained.  The difficulty with the assignment of errors in the method described so far results from the appearance of the constraint term $\lambda C$ in the Lagrangian, so that the stationary point for the constrained optimum is at a saddle point in the extended parameter space $(\lambda, \mbf{X})$.  Obviously, the solution is to devise a methodology in which no constraint appears, so that the Hessian $\tens{H}$ of the merit function $F$ at its optimal value $\mbf{G}(\mbf{X}_F) = 0$ provides an estimate of the variance of the model parameters.  As the normalization of the constraint $C$ is given by the signal energy $E_y$, that is where we will look for improvements.

Rather than constrain the spectral energy $E_\mbf{X}$ to the value given by the sum of squared data values $E_y$ in Equation~(\ref{eqn:Eysum}), what we need to do is evaluate the probability of $E_\mbf{X}$ for any set of coefficients given the data $\mbf{y}$ and its variance $\tens{V}$.  Let us rewrite the normalized signal energy in terms of the expectation value of the squared data values, \beq
E_y / N_d \Delta_t = \langle y_t^2 \rangle_{\tens{W}} \equiv \mbf{y}^T \tens{W} \mbf{y} / \tr \tens{W} \;,
\eeq here using the unnormalized weight matrix $\tens{W} \equiv \tens{V}^{-1}$.  Simply assigning that value a variance given by \beq
\langle ( y_t^2 - \langle y_t^2 \rangle_{\tens{W}} )^2 \rangle_{\tens{W}} =  \langle y_t^4 \rangle_{\tens{W}} - \langle y_t^2 \rangle_{\tens{W}}^2
\eeq turns out to be a bad idea, as the distribution for the energy $p(E_y|\mbf{y},\tens{V})$ is nothing like a Gaussian.

\begin{figure}[t]
\includegraphics[width=\columnwidth]{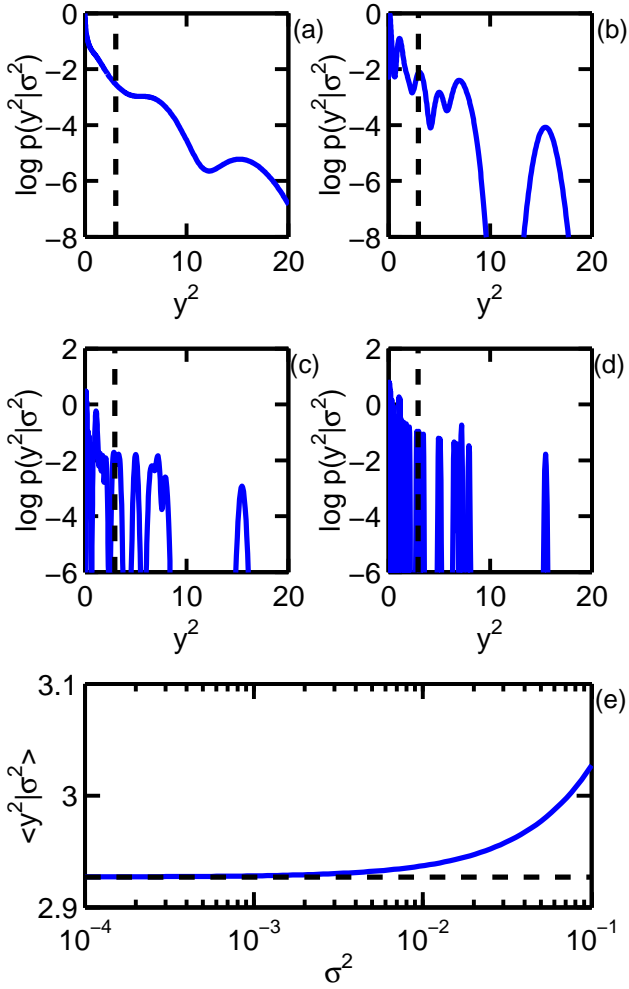}
\caption{The logarithm of the distribution $p(y^2 | \sigma^2)$ is shown for $\sigma^2 = 10^{-1}$ in (a), $10^{-2}$ in (b), $10^{-3}$ in (c), and $10^{-4}$ in (d), where the expectation value of $y^2$ is indicated by a dashed line.  The expectation value as a function of $\sigma^2$ is shown in (e), where the arithmetic mean of the squared data values is indicated by a dashed line}
\label{fig:M}       
\end{figure}

As $E_y$ and $\langle y_t^2 \rangle_{\tens{W}}$ differ only by a scaling factor $N_d \Delta_t$, let us focus our attention on the distribution $p(y^2 | \mbf{y},\tens{V})$.  Considering some signal $\mbf{y}$, first suppose that its variance is proportional to the identity matrix $\tens{V} = \sigma^2 \tens{I}$.  The individual datum distributions for $k \in [1, N_d]$ are then given by normalized Gaussians, \beq
p(y_k | y_t, \sigma^2) = (2 \pi \sigma^2)^{-1/2} \exp^{-1/2} \left[ ( y_k - y_t )^2 / \sigma^2 \right] \;,
\eeq which need to be folded over to a strictly positive axis $z_k \equiv \abs{y_k}$ by writing \bea
p(z_k | y_t, \sigma^2) &\propto& \exp^{-1/2} \left[ ( z_k - y_t )^2 / \sigma^2 \right] \\
 & & + \exp^{-1/2} \left[ ( - z_k - y_t )^2 / \sigma^2 \right] \;.
\eea  That distribution is marginalized over $k$ so that \beq
p(z | \mbf{y}, \sigma^2) = \sum_k p(z_k | y_t, \sigma^2) / N_d
\eeq may then be scaled for $y^2 = z^2$ to yield \beq
p(y^2 | \mbf{y}, \sigma^2) = p(z | \mbf{y}, \sigma^2) \abs{dz / dy^2} \;,
\eeq which gives the distribution for the estimate of the signal energy given the data and its variance.  For the signal $\mbf{y}$ in Figure~\ref{fig:E}, let us evaluate $p(y^2 | \mbf{y}, \sigma^2)$ for $\sigma^2 \in [10^{-4}, 10^{-1}]$.  In Figure~\ref{fig:M} panels (a) through (d) we show the logarithm of that distribution for the four values of $\sigma^2$ equal to powers of 10, and in panel (e) we show the expectation value $\int y^2 p(y^2 | \mbf{y}, \sigma^2) dy^2$ as a function of $\sigma^2$.  As $\sigma^2 \rightarrow 0$, the distribution in $y^2$ approaches a sum of delta distributions located at the values of $y_t^2$, and the expectation value approaches the arithmetic mean of the squared data values.

\begin{figure}[t]
\includegraphics[width=\columnwidth]{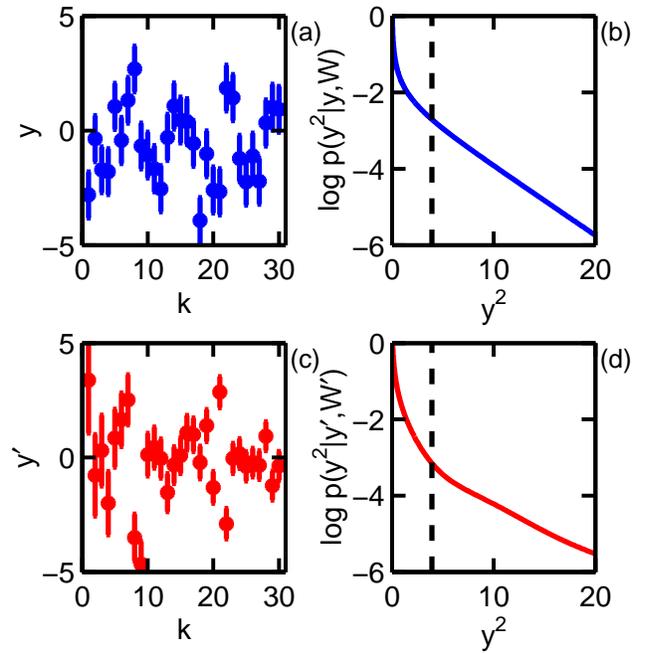}
\caption{The datum distributions indexed by $k$ in panel (a) with independent unit variance have a distribution for normalized energy shown in (b).  Assuming the data variance is given by a symmetric Toeplitz matrix with the same trace, the orthogonalized data in (c) have the distribution for normalized energy shown in (d), which has the same expectation value for $y^2$}
\label{fig:N}       
\end{figure}

If the variance matrix for the data is not proportional to the identity, one has to account for the covariance (off-diagonal elements) when evaluating the distribution for the signal energy.  To orthogonalize the data, one must decompose the weight matrix $\tens{W}$ into its eigenvalues $\tens{W}^\prime$ and eigenvectors $\tens{E}$, so that \beq
\mbf{y}^T \tens{W} \mbf{y} \rightarrow \mbf{y}^T \tens{E}^T \tens{W}^\prime \tens{E} \mbf{y} \equiv \mbf{y}^{\prime T} \tens{W}^\prime \mbf{y}^\prime
\eeq yields the same expectation value for $y^2$.  The orthogonalized datum distributions \beq
p(y_k | y_t^\prime, w_k^\prime) = (2 \pi / w_k^\prime)^{-1/2} \exp^{-1/2} \left[ ( y_k - y_t^\prime )^2 w_k^\prime \right] \;,
\eeq where $w_k^\prime \equiv W_{kk}^\prime$, are then folded, marginalized, and scaled as before to yield $p(y^2 | \mbf{y}^\prime, \tens{W}^\prime)$.  For the given signal $\mbf{y}$, we first test the method for $\tens{V} = \tens{I}$ with trace $N_d$, with the results shown in Figure~\ref{fig:N}~(a) and (b).  In this case $\mbf{y}^\prime = \mbf{y}$, and $p(y^2 | \mbf{y}^\prime, \tens{W}^\prime)$ is the same as $p(y^2 | \mbf{y}, \sigma^2 = 1)$ from the previous paragraph.  Now we suppose the data variance is given by a symmetric Toeplitz matrix $V_{jk} = 1 / (1 + \abs{j-k})$ with the same trace $N_d$.  The orthogonalized data $\mbf{y}^\prime$ shown in panel (c) have a distribution $p(y^2 | \mbf{y}^\prime, \tens{W}^\prime)$ shown in (d) with the same expectation value $\int y^2 p(y^2 | \mbf{y}^\prime, \tens{W}^\prime) dy^2$ as before.  The orthogonalization of the data does not affect the expectation value of the normalized energy distribution but does affect its shape according to the independent datum distributions.

To make contact with our goal, we must now identify $y^2$ with the normalized spectral energy $E_\mbf{X} / N_d \Delta_t$ given by the amplitude coefficients $\mbf{A}$.  The merit function $F_{RPQ} \equiv R + P + Q$ now includes a term $Q \equiv - \log p(E_\mbf{X} | \mbf{y}, \tens{V})$ for the distribution of the spectral energy and does not include any explicit constraint.  The expression $E_\mbf{X}$ is not a ``hidden'' variable but rather an auxiliary variable defined in terms of the model parameters.  The term $Q$ is in effect (the negative logarithm of) an additional prior factor in the amplitude space $\mbf{A}$ which accounts for the distribution of the spectral energy given the data and its variance.  Furthermore, the factors $E_y$ and $\mu$ must now be rewritten in terms of $E_\mbf{X}$ where they appear in the Poisson prior $P$ of Equation~(\ref{eqn:Pprior}).  The implementation of these improvements is currently under investigation; we have outlined only one possible approach here, and some variation might be found to work better in practice.

\section{Conclusions}
\label{sec:concl}
In this article we have applied the principle of maximum entropy to power spectral density estimation using the one-sided Fourier transform in the context of discrete, irregular signal sampling.  We have dispensed with the arbitrary weighting of the entropy term in the merit function in favor of a constraint on the spectral energy.  The prior is rewritten in terms of the continuous Poisson distribution whose Stirling approximation gives the familiar entropy expression for unnormalized distributions.  In the limit of vanishing errors, the spectrum with maximum evidence is equal to that with maximum likelihood given by the forward transform coefficients, and in the limit of extreme errors it approaches a flat power spectrum with the same energy as the signal.  An outline of improvements to the method to obtain the variance of the spectral coefficients is also given.

As an example, we have evaluated the power spectrum of an irregularly sampled record of stellar luminosity for the star VCas.  Several prominent peaks in the spectrum survive the smoothing action of the MaxEnt algorithm, which prevents the overestimation of structure in the spectrum when confronted with imperfect data.  In that sense, the MaxEnt algorithm gives a more conservative estimate for the power spectrum than the forward transform.  As actual measurements necessarily are accompanied by measurement errors, the incorporation of their effect through the principle of maximum entropy is suggested to those who use the Fourier transform for power spectral density estimation.

\section{Appendix}
\label{sec:app}
In deriving the usual entropy expression, recourse is made to the discrete Poisson distribution \bea
\p{k}{\mu} = e^{-\mu} \mu^k / k! \propto \mu^k / k!
\eea with parameter $\mu$ for integer $k \geq 0$, where the proportionality is given by the normalization $\sum_{k=0}^\infty \mu^k / k! = e^\mu$.  For $\mu$ and $k$ in some units except for the exponent, there are $k$ unit factors in the numerator which cancel $k$ unit factors in the denominator.  While it has been suggested~\citep{CaMwA-1187} that the cumulative distribution is what should be generalized to the continuum, it seems more intuitive that the probability density be generalized through the obvious substitution $k! \rightarrow \Gamma(n+1)$ for continuous $n \geq 0$, giving a continuous Poisson density $\p{n}{\mu} \propto \mu^n / \Gamma(n+1)$.  About the only objection one could raise for such a density function is one of normalization over an infinite axis, which itself is mooted by consideration of some finite cutoff $n_\mrm{max}$ larger than any scale of interest in the problem at hand.  The use of an unnormalizable distribution is implicit in the maximum likelihood method, as the uniform distribution is only finite when considered over a restricted domain.

As almost any non-negative function with finite domain can be normalized to a probability density, let us consider the normalization of the continuous Poisson distribution over an infinite axis.  Can we show that $\int_0^\infty dn\, \mu^n / \Gamma(n+1) = e^\mu$?  Taking the derivative with respect to the parameter $\mu$ of both sides, the RHS is the definition of the exponential function, $\dsub{\mu} e^\mu \equiv e^\mu$.  The derivative of the LHS leads to the expression \bea
\dsub{\mu} \int_0^\infty \dfrac{\mu^n \,dn}{\Gamma(n+1)} = \int_0^\infty \dfrac{n \mu^{n-1} \,dn}{\Gamma(n+1)} = \int_0^\infty \dfrac{\mu^{n-1} \,dn}{\Gamma(n)} \;,
\eea whereupon shifting the limits down by one unit, \bea
\int_{-1}^\infty \dfrac{\mu^n \,dn}{\Gamma(n+1)} &=& \int_{-1}^0 \dfrac{\mu^n \,dn}{\Gamma(n+1)} + \int_0^\infty \dfrac{\mu^n \,dn}{\Gamma(n+1)} \\
 &=& \int_0^\infty \dfrac{\mu^n \,dn}{\Gamma(n+1)} \;,
\eea where the last step follows from the observation that our original object, the probability density, is zero over the negative axis, $\p{n}{\mu} = 0$ for $n<0$.  We hesitate to call this example a proof of the relation, as the most formal of mathematicians might object to the heuristic final step, but it is certainly highly suggestive that the normalization carries over to the continuum unaltered.


%
\acknowledgments
We acknowledge with thanks the variable star observations from the AAVSO International Database contributed by observers worldwide and used in this research.

%
\bibliographystyle{spr-mp-nameyear-cnd}  

\end{document}